\documentclass[conference]{IEEEtran}
\IEEEoverridecommandlockouts
\usepackage{cite}
\usepackage{amsmath,amssymb,amsfonts}
\usepackage{algorithmic}
\usepackage{graphicx}
\usepackage{textcomp}
\usepackage{xcolor}
\usepackage{url}
\def\BibTeX{{\rm B\kern-.05em{\sc i\kern-.025em b}\kern-.08em
    T\kern-.1667em\lower.7ex\hbox{E}\kern-.125emX}}
\begin{document}

\title{CautionSuicide: A Deep Learning Based Approach for Detecting Suicidal Ideation in Real Time Chatbot Conversation\\
}

\author{\IEEEauthorblockN{Nelly Elsayed}
\IEEEauthorblockA{\textit{School of Information Technology} \\
\textit{University of Cincinnati}\\
Ohio, United States\\
elsayeny@ucmail.uc.edu}
\and
\IEEEauthorblockN{Zag ElSayed}
\IEEEauthorblockA{\textit{School of Information Technology} \\
\textit{University of Cincinnati}\\
Ohio, United States \\
elsayezs@ucmail.uc.edu}
\and
\IEEEauthorblockN{Murat Ozer}
\IEEEauthorblockA{\textit{School of Information Technology} \\
\textit{University of Cincinnati}\\
Ohio, United States\\
ozermm@ucmail.uc.edu}
}

\maketitle

\begin{abstract}
Suicide is recognized as one of the most serious concerns in the modern society. Suicide causes tragedy that affects countries, communities, and families. There are many factors that lead to suicidal ideations. Early detection of suicidal ideations can help to prevent suicide occurrence by providing the victim with the required professional support, especially when the victim does not recognize the danger of having suicidal ideations. As technology usage has increased, people share and express their ideations digitally via social media, chatbots, and other digital platforms. In this paper, we proposed a novel, simple deep learning-based model to detect suicidal ideations in digital content, mainly focusing on chatbots as the primary data source. In addition, we provide a framework that employs the proposed suicide detection integration with a chatbot-based support system.
\end{abstract}


\section{Introduction}

\begin{figure}[htbp]
	\centerline{\includegraphics[width=8.5cm, height= 4 cm]{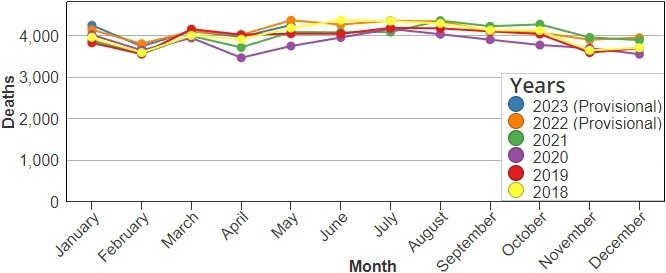}}
	\caption{The number of committed suicides each month from 2018 to 2023 according to the Center for Disease Contol and Prevention (CDC) data published in~\cite{cdc_statistics2}}.
	\label{suicide_rates}
\end{figure}

Suicide is an intentional act of an individual to cause one's own death~\cite{turecki2019suicide}. According to the Centers for Disease Control and Prevention (CDC) Data and Statistics Fatal Injury Report for 2021, suicide is the 11$^{\mathrm{th}}$ leading cause of death in the United States~\cite{cdc_statistics}. There are many factors that can increase the risk of a suicide attempt. Mental disorders such as depression and alcohol use disorders can lead to suicide~\cite{chesney2014risks}. Suicide can be connected to other injury and violence forms, such as experiencing abuse, bullying, cyberbullying, sexual violence, and other types of violence~\cite{rivara2019effects,azumah2023cyberbullying,hinduja2010bullying,hinduja2019connecting}. In addition, suicide has been found amongst vulnerable groups who experience discrimination and social abuse, such as refugees, migrants, indigenous, prisoners, and individuals under supervision period~\cite{who_suicide,nih_report}. Figure~\ref{suicide_rates} shows the number of committed suicides each month from 2018 to 2023 according to the Centers for Disease Control and Prevention (CDC)~\cite{cdc_statistics2}.

Preventing suicide requires collaboration and coordination between different social sectors, including education, health, labor, law enforcement, defense, media, and business~\cite{stone2017preventing,world2004promoting}. Integrated efforts can provide a significant supportive approach for people, especially those at high risk for suicide action~\cite{mann2005suicide}. 

Suicidal ideation detection is one of the crucial steps to prevent suicide. Early detection of suicidal ideations can lead to saving lives and provide the needed medical and emotional support for individuals who may perform suicide~\cite{al2021suicidal,cole2013suicide}. There are several domains where suicidal ideations can be investigated, including clinical-based and non-clinical-based schemes~\cite {devylder2015psychotic,ji2020suicidal}. The clinical-based ones are categorized as the clinical interviews that are performed by medical professionals, questionnaires, electronic health records, and suicide notes that are performed within the clinical stay. The non-clinical is based on social content (posts and comments) and suicide notes.  

Due to the wide usage of technology and digital plattforms, even several clinical visits have been transformed to virtual environment~\cite{wosik2020telehealth,mcgrail2017virtual,mann2020covid,rizzo2002virtual}. Thus, the virtual chat including chatbots have been widley increased in different busieness, industry, and medical sectors~\cite{weissensteiner2018chatbots,luo2022critical,skrebeca2021modern,elsayed2023machine}. Unitlizing such chatbot systems to help detecting suicidal ideation is curcial to save lives~\cite{fonseka2019utility,martinengo2019suicide}.

Thus, in this paper, we proposed a novel, simple deep learning model to detect suicide ideation and a chatbot framework that can utilize such a model for early detection of suicide ideation. The contributions of this paper can be summarized as follows:
\begin{itemize}
	\item A novel simple deep gate recurrent unit-based model to detect suicidal ideation in digital text content.
	\item A chatbot-based framework that utilizes the proposed model to investigate and detect suicide ideation for users of the chatbot. In addition, the framework can provide additional reports for medical providers or corresponding authorities based on the issuer and the chatbot's purpose.
	\item Provide open questions and limitations researchers can address to employ the proposed framework for different applications and targets. 
\end{itemize}

\begin{figure*}[htbp]
	\centerline{\includegraphics[width=12cm, height= 7 cm]{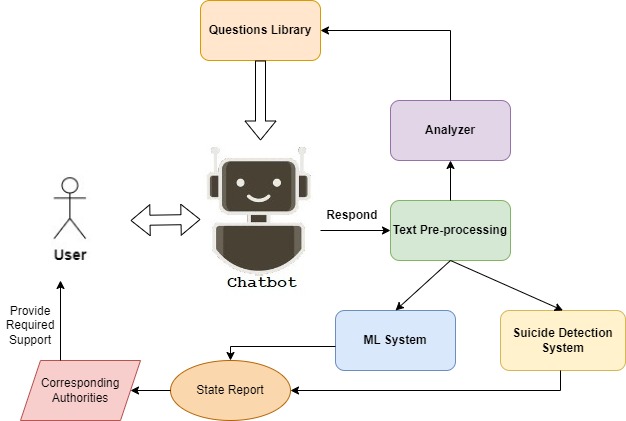}}
	\caption{The proposed framework for suicidal ideation detection in a chatbot conversationn.}
	\label{framework}
\end{figure*}

\section{Suicide Detection in Chatbot Conversation Framework}
The proposed framework for detecting suicidal ideations in chatbot conversation is shown in Figure~\ref{framework}. The proposed framework aims to integrate the suicidal thought detections for offenders under correction~\cite{elsayed2023machine}. First, the user starts to chat with the chatbot by entering text input via a friendly graphical user interface (GUI). The questions that the chatbot will provide to the user are based on previously prepared questions by psychology specialists and address the major conditions that the corresponding authorities looking for ad the major purpose of the chatbot (e.g., if the corresponding authorities can be a medical provider if the chatbot system based on detecting specific mental disorders and corresponding authorities can be the related to rehabilitation center when it address individuals under the correction period). The selection of the following questions is based on the answers provided by the user after text preprocessing and keyword analysis. The preprocessed text will also be provided to the suicide detection system  to detect suicide ideations and to the machine learning system to detect other mental health issues. Then, a final report, including the overall findings, will be provided to the corresponding authorities to perform actions and provide the required support to the individual. The proposed framework can be directly employed in the digital correction monitoring support system as well as in other related chatbot applications~\cite{elsayed2023machine}.

\begin{figure}[htbp]
	\centerline{\includegraphics[width=7cm, height= 10 cm]{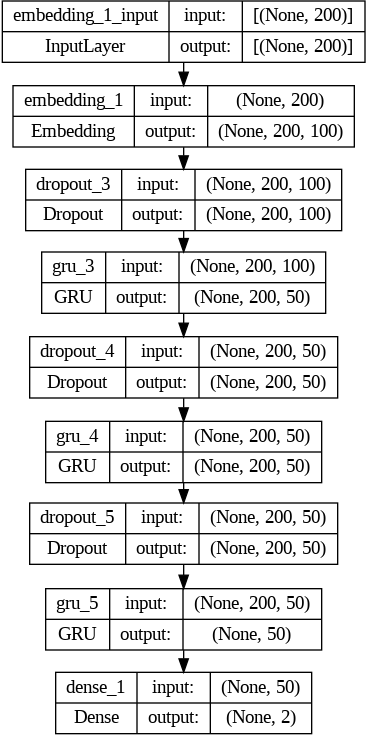}}
	\caption{The proposed model for the suicidal ideation detection diagram.}
	\label{model}
\end{figure}

\begin{table}[b]
	\caption{The proposed model layers and their corresponding number of parameters.}
	\begin{center}
		\begin{tabular}{|l|c|}
			\hline
			\textbf{Layer}& \textbf{\#Param}\\
			\hline
			Embedding& 1000000     \\
			Dropout& 0\\
			GRU&22800     \\
			Dropout&0\\
			GRU&15300     \\
			Dropout&0 \\
			GRU&15300     \\
			Dense&102       \\
			Total param&1053502 \\
			Trainable params&53502 \\
			Non-trainable params&1000000 \\
			\hline
		\end{tabular}
		\label{table_summary}
	\end{center}
\end{table}

\section{Suicide ideation Detection Model}
The proposed suicide ideation detection model diagram is shown in Figure~\ref{model}. The gated recurrent unit (GRU) has been selected as the main component of suicide detection. The selection has been set based on the efficiency of GRU in learning short-term dependencies compared to other recurrent architectures such as the long short-term memory (LSTM)~\cite{chung2014empirical,elsayed2018deep,wang2018gated}. In addition, the GRU has a simple architecture and can be trained faster than the LSTM~\cite{chung2014empirical}. Moreover, based on the current analysis and proposed models for suicide detection, the GRU has not been implemented as a standalone architecture for this purpose~\cite{haque2022comparative}. Table~\ref{table_summary} shows a summary of the model layers and their corresponding number of parameters. The non-trainable parameters are corresponding to the embedding layer paramenters~\cite{wu2019stochastic}.

\section{Experiment and Results}

\subsection{Dataset}
In our experiments, we used the suicide ideation dataset on Kaggle~\cite{dataset}. The dataset consists of a collection of posts from subreddits of the Reddit platform, including "SuicideWatch" and "Depression" subreddits. The collected posts from SuicideWatch are labeled as suicide. The collected posts from the depression subreddit are labeled as depression. The non-suicide posts are collected from r/teenagers. The dataset consists of 233,337 data samples, where 117,301 are labeled as suicide and 116,039 are labeled as non-suicide. 

The data has been preprocessed, including removing stopwords, punctuations, emails, HTML tags, special characters, and accented characters. Then, the data was tokenized using Keras text preprocessing tokenizer, setting the number of words to 100,000. Then, the tokenized text has been converted to numeric sequences with post padding~\cite{alam2019impact,vijayarani2015preprocessing}. Then, for model training and testing, the data labels have been coded as class 0 and class 1 for suicide and non-suicide, respectively. The data has been split to train, validate, and test the dataset with ratios 50\%, 20\%, and 30\%, respectively.

\begin{figure}[htbp]
	\centerline{\includegraphics[width=7cm, height= 3 cm]{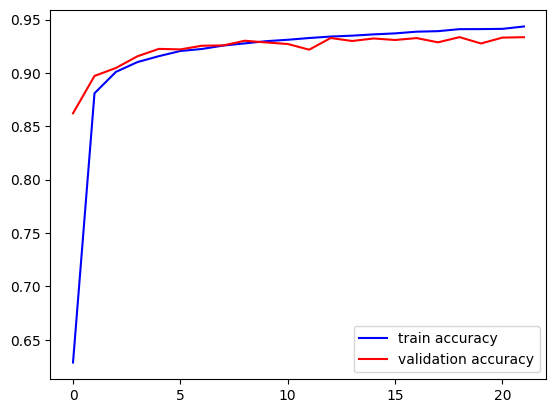}}
	\caption{The training versus validation accuracies of the proposed suicide ideations detection model.}
	\label{accuracy}
\end{figure}

\begin{figure}[htbp]
	\centerline{\includegraphics[width=7cm, height= 3 cm]{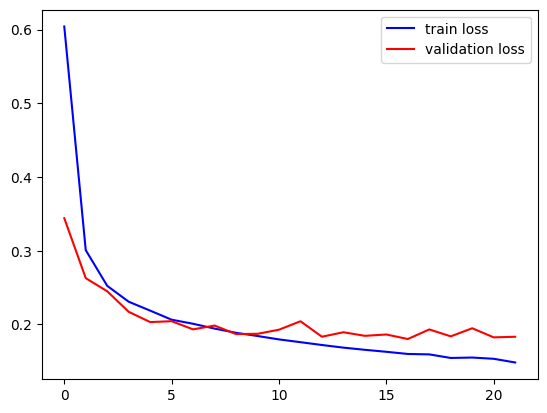}}
	\caption{The training versus validation losses of the proposed suicide ideations detection model.}
	\label{loss}
\end{figure}

\begin{table}[htbp]
	\caption{The overall statistics of the testing process of the proposed suicide ideation detection model.}
	\begin{center}
		\begin{tabular}{|l|c|}
			\hline
			\textbf{Merit}& \textbf{Value}\\
			\hline
			95\% CI                                     &                        (0.93877,0.94783)\\
			Train Accuracy                                   &                  95.248\%     \\
			Test Accuracy                                   &                 94.330\% \\     
			F1 Macro                                   &                       0.94330\\
			F1 Micro                                   &                       0.94330\\
			Hamming Loss                                &                      0.0567\\
			Reference Entropy                           &                      0.99989\\
			Response Entropy                             &                     1.0\\
			Standard Error                               &                     0.00231\\
			Kappa                                         &                   0.8866\\
			Kappa Standard Error                          &                    0.00463\\
			SOA1(Landis \& Koch)                          &                    Almost Perfect\\
			SOA2(Fleiss)                                  &                    Excellent\\
			SOA3(Altman)                                  &                    Very Good\\
			\hline
		\end{tabular}
		\label{tab2}
	\end{center}
\end{table}

\subsection{Experimental Setup}

The proposed model has been implemented using Python 3.10.12 and Tensorflow 2.15.0. We used Google Colaboratory (Colab) to write and execute our experiments. The proposed model embedding dimension has been set to 100 and the dropout rates to 20\%. The softmax function has been used as the activation of the dense layer~\cite{goodfellow2016deep}. The GRU number of unrollments is set to 50, with  glorot\_uniform and orthogonal as the kernel and recurrent initializers, respectively~\cite{glorot2010understanding,vorontsov2017orthogonality}. The number of epochs is set to 25, with early stopping occurring in 22 epochs. The batch size has been set to 128. The categorical cross-entropy is set as the loss function. The optimizer is set to Adam optimizer~\cite{kingma2014adam}.

The proposed model training versus validation accuracy and loss through the training epochs are shown in Figure~\ref{accuracy} and Figure~\ref{loss}, respectively. Table~\ref{tab2} shows the overall statistics of the proposed model where 95\% CI indicates the 95\% confidence interval, and SOA is the strength of agreement. 

\begin{figure}[htbp]
	\centerline{\includegraphics[width=4cm, height= 4 cm]{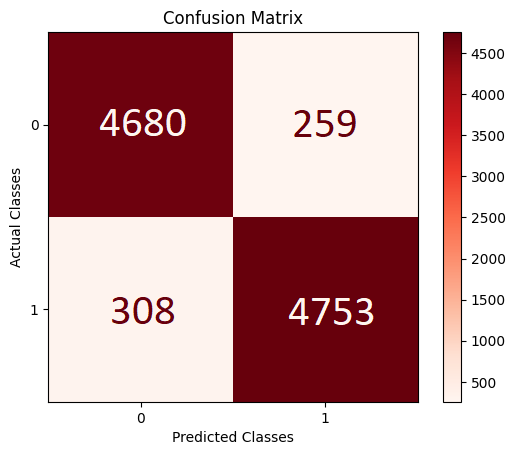}}
	\caption{The confusion matrix of testing the proposed suicidal ideation detection model.}
	\label{confusion_matrix}
\end{figure}

On the classes-level (suicide and non-suicide) model statistics, Table~\ref{tab1} shows the class statistics of the testing process for an average of (n=3) trials of the proposed model where FNR and FPR are the false negative rate and false positive rate, respectively. Figure~\ref{confusion_matrix} shows the confusion matrix of testing the proposed model.

\begin{table}[htbp]
	\caption{The class statistics of the testing process of the proposed suicide ideation detection model.}
	\begin{center}
		\begin{tabular}{|l|c|c|}
			\hline
			\textbf{Merit}& \textbf{Suicide}& \textbf{Non-Suicide} \\
			\hline
		     AGF(Adjusted F-score)                                       &     0.94608    &   0.94053  \\  
		     AUC(Area under the ROC curve)                              &      0.94335    &   0.94335     \\
		     ERR(Error rate)                                             &        0.05670     &   0.05670   \\
		     FNR & 0.05244    &   0.06086\\
		     FPR &                            0.06086    &   0.05244     \\  
		     
		     Sensitivity   &     0.94756    &   0.93914   \\
		     Specificity             &    0.93914  &     0.94756     \\
		     Precision &  0.93825    &   0.94832        \\
		     F1-Score &     0.94288    &   0.94371  \\
		     
		     		\hline
		\end{tabular}
		\label{tab1}
	\end{center}
\end{table}

Comparing the proposed model to the existing state-of-the-art models that have been analyzed in~\cite{haque2022comparative}. Table~\ref{compare} shows a comparison between the proposed model and the state-of-the-art models for suicide ideation detection in text. The proposed model exceeds the state-of-the-art models in overall accuracy, precision, recall, F1-score, and area under the curve (AUC). Thus, the proposed model is promising to provide precise accuracy in detecting suicide ideations in text, which can help to save lives.

\begin{table}[htbp]
	\caption{A comparison between the proposed suicide ideation detection model and other state-of-the-art models.}
	\begin{center}
		\begin{tabular}{|l|c|c|c|c|c|}
			\hline
			\textbf{Model}& \textbf{Accuracy}& \textbf{Precision} & \textbf{Recall} &  \textbf{F1-Score} &\textbf{AUC} \\
			\hline
			
			RF &93.0\% &0.92& 0.92 &0.92& 0.92\\
			SVC& 91.9\% &0.91& 0.91 &0.91 &0.91\\
			SGD &91.7\% &0.91 &0.91 &0.91 &0.91\\
			LR &91.2\% &0.91 &0.91& 0.91 &0.91\\
			MNB &84.6\% &0.84 &0.84& 0.84 &0.84\\
			BiLSTM& 93.6\% &0.93 &0.93 &0.93 &0.93\\
			LSTM &93.5\%& 0.93& 0.93& 0.93& 0.93\\
			BiGRU &93.4\% &0.93 &0.93& 0.93 &0.93\\
			CLSTM &93.2\% &0.91& 0.93 &0.93& 0.93\\

			\textbf{GRU (our)}& \textbf{94.33\%}& \textbf{ 0.9433} & \textbf{ 0.9433}  & \textbf{0.9433}  &  \textbf{ 0.9433} \\
			
			\hline
		\end{tabular}
		\label{compare}
	\end{center}
\end{table}

\section{Limitations and Open Questions}
The limitations of the proposed framework is that the privacy and security aspects of data and user identifications must be addressed as two major components. Thus, we recommend researcher and who will deploy such systems to investigate and address these aspects while implementing the framework.

\section{Conclusion}
Detecting suicide ideation can help to save lives by providing the required support in a timely manner. Thus, contributing toward saving human lives, in this paper, we proposed a novel, simple model based on the gated recurrent units to detect suicidal ideation in digital text content. In addition, we proposed a chatbot framework that can utilize the proposed suicide ideation detection model, which can be used in different fields such as education, medicine, and law enforcement. The proposed model exceeds the state-of-the-art models in detecting suicide ideations, and it can be simply replicated, which can significantly help to save lives and prevent suicide.



\bibliographystyle{ieeetr}
\bibliography{references_suicide}

\end{document}